\documentstyle[aps,twocolumn,prl,epsfig]{revtex}

\newcommand{\be}{\begin{equation}}
\newcommand{\ee}{\end{equation}}
\newcommand{\bea}{\begin{eqnarray}}
\newcommand{\eea}{\end{eqnarray}}
\newcommand{\mel}[3]{<\!#1\,|\,#2\,|\,#3\!>}

\newcommand{\scp}[2]{\mbox{$<\!#1\,|\,#2\!>$}}

\newcommand{\ket}[1]{\mbox{$|\,#1\!>$}}

\begin{document}

\tighten\draft
%\draft
%\preprint{}

\twocolumn[\hsize\textwidth\columnwidth\hsize\csname@twocolumnfalse\endcsname

%\title{Inverse Hartree--Fock Theory}
\title{Hartree--Fock Approximation for Inverse Many--Body Problems}
\author{J. C.  Lemm and J. Uhlig} 
\address{
Institut f\"ur Theoretische Physik I,
Universit\"at M\"unster, 48149 M\"unster, Germany
}
\date{\today}
\maketitle

\begin{abstract}
A new method is presented to reconstruct
the potential of a quantum mechanical many--body system
from observational data,
combining a nonparametric Bayesian approach
with a Hartree--Fock approximation.
A priori information 
is implemented 
as a stochastic process, 
defined on the space of potentials.
The method is computationally feasible
and provides a general framework to treat
inverse problems for quantum mechanical many--body systems.
\end{abstract}
%21.60.Jz Hartree-Fock and random-phase approximation
%02.50.Rj Nonparametric inference
%02.50.Wp Inference from stochastic processes
] %end bracket for twocolumn

\narrowtext

The reconstruction of inter--particle forces
from observational data
is of key importance for any application of quantum mechanics
to real world systems.
Such inverse problems have been studied intensively 
in inverse scattering theory 
and in inverse spectral theory
for one--body systems in one and, later, in three dimensions
\cite{iqs:Newton:1989,iqs:Chadan-Colton-Paivarinta-Rundell:1997}.
In this Paper we now outline a method,  
designed to deal with inverse problems for many--body systems.

Being the mathematical counterpart
of induction problems in philosophy,
inverse problems appear quite naturally in science
when justification of a physical law
has to be based on a finite number of observations.
Such problems are notoriously ill--posed
in the sense of Hadamard \cite{iqs:Tikhonov-Arsenin:1977,iqs:Kirsch:1996}.
In that case it is well known that 
additional a priori information
is required to obtain a stable and unique solution.
Referring to a Bayesian framework
\cite{iqs:Berger:1980,iqs:Robert:1994,iqs:Gelman-Carlin-Stern-Rubin-1995},
we implement a priori information
in form of stochastic processes over potentials
\cite{iqs:Lemm:1999b}.
While a standard procedure is 
to fit parameterized potentials to the data,
we will especially be interested
in less restrictive, nonparametric approaches.
As, up to now, calculating an exact solution 
of inverse many--body problems is not feasible,
we will treat the problem 
in an `inverse Hartree--Fock approximation' (IHFA).

A main advantage of Bayesian methods is their flexibility.
They can easily be adapted to different learning situations
and have therefore been applied 
to a variety of empirical learning problems,
including classification, regression, density estimation
\cite{iqs:Neal:1996,iqs:Williams-Rasmussen:1996},
and, recently, to quantum statistics \cite{iqs:Lemm:1999b}.
In particular, within a Bayesian approach it is straightforward  
to deal with measurements of arbitrary quantum mechanical observables, 
to include classical noise processes,
and to implement a priori information
explicitly in terms of the potential.
Computationally, on the other hand,
working with stochastic processes, 
or discretized versions thereof,
is much more demanding
than, for example, fitting parameters.
This holds especially for applications to quantum mechanics
where one can not take full advantage of the convenient analytical features
of Gaussian processes.
Due to increasing computational resources, however,
the corresponding learning algorithms become now
numerically feasible.

We will consider many--fermion systems with Hamiltonians, 
$H$ = $T+V$, consisting
of a one--body part $T$,
e.g., $T$ = $-(1/2m)\Delta$ (with Laplacian $\Delta$,
mass $m$, $\hbar$ = 1),
and a two--body potential $V$.
To write such Hamiltonians in second quantization,
we introduce creation and annihilation operators
$a_\alpha^\dagger$, $a_\alpha$ corresponding to a complete 
single particle basis $\ket{\varphi_\alpha}$,
i.e.,\ 
$a_\alpha^\dagger\ket{0}$ = $\ket{\varphi_\alpha}$
and
$a_\alpha\ket{0}$ = 0.
As we are dealing with fermions,
we have to require the usual anticommutation relations
$a_\alpha a_\gamma^\dagger +a_\gamma^\dagger a_\alpha $ 
= 
$\scp{\varphi_\alpha}{\varphi_\gamma}$, 
$a_\alpha^\dagger a_\gamma^\dagger +a_\gamma^\dagger a_\alpha^\dagger $ 
=0,
$a_\alpha a_\gamma +a_\gamma a_\alpha $ 
=0,
and we can write
\begin{equation}
H 
= \sum_{\alpha\beta}T_{\alpha\beta} \,a^\dagger_\alpha a_\beta
+ \frac{1}{4}\sum_{\alpha\beta\gamma\delta} V_{\alpha\beta\gamma\delta} 
  \, a^\dagger_\alpha a^\dagger_\beta a_\delta a_\gamma 
,
\end{equation}
with 
$T_{\alpha,\beta}$ = $<\!\varphi_\alpha|T|\varphi_\beta\!>$
and antisymmetrized matrix elements 
$V_{\alpha\beta\gamma\delta}$
%=$-V_{\alpha\beta\delta\gamma}$
%=$-V_{\beta\alpha\gamma\delta}$
=
\mbox{$<\!\varphi_\alpha\varphi_\beta|V|\varphi_\gamma\varphi_\delta\!>$}.
We will consider two--body potentials %$V$
which are local and depend only on
the distance  
between the particles $x$ = $|x_1-x_2|$, 
i.e., 
$V_{x_1 x_2 x^\prime_1 x^\prime_2}$
=
$v(|x_1-x_2|) \big(
\delta(x_1-x_1^\prime) \delta(x_2-x_2^\prime)
-
\delta(x_1-x_2^\prime) \delta(x_2-x_1^\prime)\big)$.
For unknown function $v(x)$, 
our aim will be to reconstruct this function
from observational data.

To obtain information about the potential,
the system has to be prepared in a state depending on $v$.
Such a state can be a stationary statistical state, 
e.g.\ a canonical ensemble,
or a time--dependent state evolving according to the Hamiltonian
of the system.
In the following we will study many--body systems 
being prepared in their ground state.
The (normalized) $N$--particle ground state wave function $\psi_0$ 
depends on $v$ and is antisymmetrized for fermions.
As observational data
we choose $n$ simultaneous measurements of the coordinates
of the $N$ particles,
the corresponding observable being the
coordinate operator $\hat x$.
The $i$th measurement results hereby in a vector $x_i$,
consisting of $N$ components $x_{i,j}$, 
each representing a single particle coordinate.
Introducing the Slater determinant 
$\ket{x_i}$ = $\ket{x_{i,1},\cdots, x_{i,n}}$,
made of orthonormal single particle orbitals $\ket{x_{i,j}}$,
the probability density of measuring 
the coordinate vector $x_i$ given $v$ is,
according to the axioms of quantum mechanics,
\be
p(x_i|\hat x,v) 
=<\!\psi_0|x_i\!><\!x_i|\psi_0\!>
= |\psi_0(x_{i,1},\cdots, x_{i,N})|^2
,
\label{likelihood}
\ee
which, when regarded as function of $v$, for fixed $x_i$, 
is also called the likelihood of $v$.
In contrast to an ideal measurement of a classical system,
the state of a quantum system is typically
changed by the measurement process.
In particular, its state is projected 
in the space of eigenfunctions of the measured observable
with eigenvalue equal to the measurement result.
Hence, if we want to deal with 
independent, identically distributed data,
the system must  
be prepared in the same state before each measurement.
Under such conditions the total likelihood factorizes
\begin{equation}
p(x_1,\cdots,x_n|\hat x,v)
= 
\prod_{i=1}^n p(x_i|\hat x,v)
.
\label{total-likelihood}
\end{equation}

In maximum likelihood approximation,
a potential is reconstructed 
by selecting a space of parameterized potentials  $v(x,\xi)$
and maximizing the total likelihood (\ref{total-likelihood})
with respect to the parameters 
$\xi$,
i.e.,\
$v^*(x)$ = $v(x,\xi^*)$ with $\xi^*$ 
= ${\rm argmax}_{\xi} \, p(x_1,\cdots,x_n|\hat x,v(\xi))$.
This, however,
only yields a unique solution
if the parameterized space is small enough.
%as soon as the parameterized space becomes too flexible.
%Hence, if we do not want to restrict the potentials 
%to a relatively small parameterized space,
Otherwise,
additional constraints have to be included to
determine a potential uniquely.
In a Bayesian framework those constraints 
are provided by additional a priori information
and implemented by selecting a prior density $p(v)$,
interpreted as probability density 
before having received data.
The posterior density $p(v|D)$, being
the probability density of $v$ after having received data $D$, 
is then obtained according to Bayes' rule,
\begin{equation}
p(v|D)=\frac{p(v)\prod_i p(x_i|\hat x,v)}{p(x_1,\cdots,x_n|\hat x)}
.
\label{bayes-rule}
\end{equation}
Finally, the predictive density is obtained from the posterior density
as posterior expectation of the likelihood
$p(x|\hat x,D)$
= $\int \!dv\, p(x|\hat x,v)\, p(v|D)$.
Within a nonparametric approach, 
where function values $v(x)$, and not parameters,
represent the fundamental variables,
$p(v)$ and $p(v|D)$ are a stochastic processes 
and the integration over $v$ 
is a functional integration over functions $v(x)$.
Such integrations may be approximated 
by Monte--Carlo methods,
or,
as we will do in the following,
be treated in maximum a posteriori approximation, 
a variant of the saddle point method.
In that case one assumes
that the main contribution to the integral
comes from the potential with maximal posterior, i.e.,
$p(x|\hat x,D) \approx p(x|\hat x,v^*)$ 
where $v^*$ = ${\rm argmax}_v p(v|D)$.
So we are left with maximizing the posterior (\ref{bayes-rule}),
which we will do by setting the
functional derivative of the posterior 
(\ref{bayes-rule})
with respect to $v(x)$
to zero
(for $x\ge 0$).
Hence, introducing the notation
$\delta_{v(x)}$
=
$\delta/\delta v(x)$, 
we have to solve the stationarity equation
\begin{equation}
0
=\delta_{v(x)} \ln p(v)+\sum_i \delta_{v(x)} \ln p(x_i|\hat x,v)
.
\label{posterior-statEQ}
\end{equation}

The most commonly used prior processes are Gaussian.
Being technically only slightly more complicated 
but much more general,
we can also consider mixtures of Gaussian processes
\cite{iqs:Lemm:1999},
for which 
$p(v)$ = $\sum_k p(k) p(v|k)$
with Gaussian components
\begin{equation}
p(v|k) = 
\left(\det \frac{{\bf K}_k}{2\pi}\right)^\frac{1}{2}
e^{-\frac{1}{2} < v-v_k | {\bf K}_k | v-v_k >}
,
\label{gaussprior}
\end{equation}
positive (semi--)definite covariance
${{\bf K}_k}^{-1}$, 
and regression function $v_k(x)$,
playing the role of a reference potential.
A typical choice for the inverse covariance 
is the negative Laplacian multiplied with a `regularization parameter' 
$\lambda$, i.e.,
${{\bf K}_k}$ = $-\lambda \Delta$, 
which favors smooth potentials.
Higher order differential operators may be included in ${\bf K}_k$,
as it is often done in regression problems
to get differentiable regression functions
\cite{iqs:Wahba:1990,iqs:Girosi-Jones-Poggio:1995}.
Also useful are integral operators,
for example, to enforce approximate
periodicity of $v$ \cite{iqs:Lemm:1999b}.
The functional derivative 
of a Gaussian mixture prior with respect to $v$
is easily found as
\begin{equation}
\delta_{v(x)} \ln p (v) 
= -{\bf K}_0(v-v_0)
,
\label{prior-dev}
\end{equation}
where ${\bf K}_0$  =
$\sum_k p(k|v)\,{\bf K}_k$
and 
$v_0$ = 
${\bf K}_0^{-1} \sum_k p(k|v)\, {\bf K}_k v_k$
with $p(k|v)$ = ${p(v|k)p(k)}/{p(v)}$.

To calculate the functional derivative
of the likelihood (\ref{likelihood}),
$\delta_{v(x)} p(x_i|\hat x,v)$
=$<\!\delta_{v(x)} \psi_0|x_i\!><\!x_i|\psi_0\!>$
+
$<\!\psi_0|x_i\!><\!x_i|\delta_{v(x)} \psi_0\!>$,
we need $\delta_{v(x)} \psi_0$.
It is straightforward to show, 
by taking the functional derivative of 
$H\psi_0$ = $E_0\psi_0$, 
that,
for nondegenerate ground state,
a complete basis of eigenstates $\psi_\gamma$ with energies $E_\gamma$,
and 
requiring 
$<\! \psi_0 |\delta_{v(x)} \psi_0\!>$ = 0
to fix norm and phase,
\be
|\delta_{v(x)} \psi_0\!>
=\sum_{\gamma\ne 0}
\frac{1}{E_0-E_\gamma} 
|\psi_\gamma\!><\!\psi_\gamma|\delta_{v(x)} H
\ket{\psi_0}
.
\label{many-body-delta-psi}
\ee
Furthermore, from 
$\delta_{v(x)} v(|x_1-x_2|)$ = $\delta (x-|x_1-x_2|)$
directly follows
\be
\delta_{v(x)} H
=
\frac{1}{2}\sum_{x_1}  
a^\dagger_{x_1} (a^\dagger_{x_1-x} a_{x_1-x} +
a^\dagger_{x_1+x} a_{x_1+x} )a_{x_1}
.
\ee
Typically,
a direct solution of
the many--body equation (\ref{many-body-delta-psi})
is not feasible.
To get a solvable problem
we treat the many--body system
in Hartree--Fock approximation 
\cite{iqs:Eisenberg-Greiner:1972,iqs:Ring-Schuck:1980,iqs:Blaizot-Ripka:1986}
(For non--hermitian $H$ see \cite{iqs:Lemm:1995a}).
Thus,  
as first step of an IHFA,
we approximate
the many--body Hamiltonian $H$
by a one--body Hamiltonian $h$
defined self-consistently by
\be
h_{xx^\prime} 
= 
T_{xx^\prime}  
+ \sum_{k=1}^N \mel{x \varphi_k}{V}{x^\prime \varphi_k}
\label{h-hf}
,
\ee
$\varphi_k$ being the $N$--lowest 
orthonormalized eigenstates of $h$, i.e., 
\be
h\varphi_k = \epsilon_k \varphi_k, 
\qquad \epsilon_1\le \epsilon_k\le \epsilon_N
\label{HF-eq}
.
\ee
The corresponding normalized
$N$--particle Hartree--Fock ground state 
$\Phi_0$ 
= $|\varphi_1,\cdots,\varphi_N\!>$
is the Slater determinant
made of the $N$--lowest single particle orbitals $\varphi_k$.
The Hartree--Fock likelihood, replacing (\ref{likelihood}), becomes,
\be
p_{\rm HF}(x_i|\hat x,v)=
\scp{\Phi_0}{x_i}\scp{x_i}{\Phi_0}
.
\ee
To find the functional derivative 
$\delta_{v(x)} p_{\rm HF}(x_i|\hat x,v)$
= 
$\scp{\delta_{v(x)}\Phi_0}{x_i} \scp{x_i}{\Phi_0}$
+
$\scp{\Phi_0}{x_i} \scp{x_i}{\delta_{v(x)}\Phi_0}$,
we define the overlap matrix $B_i$ 
with matrix elements
$B_{kl;i}$ 
= $\varphi_k(x_{i,l})$ 
= $\scp{x_{i,l}}{\varphi_k}$
and expand 
$\scp{x_i} {\Phi_0}$
= $\det B_i$
= $\sum_l^N M_{kl;i} B_{kl;i}$,
in terms of its cofactors 
$M_{kl,i}$ = $(B_i^{-1})_{lk}\det B_i$.
Applying the product rule yields
\be
\delta_{v(x)} \scp{x_i}{\Phi_0}
= \sum_{kl}^N M_{kl;i} \, \delta_{v(x)} {\varphi_k}(x_{i,l})
.
\ee

Again, we proceed by taking the functional derivative of
Eq.\ (\ref{HF-eq})
and obtain after standard manipulations
(for nondegenerate $\epsilon_k$
and $<\!\varphi_k|\delta_{v(x)} \varphi_k\!>$ = 0),
\be
|\delta_{v(x)} \varphi_k>
=
\sum_{k\ne l} \frac{1}{\epsilon_k-\epsilon_l}
|\varphi_l><\varphi_l|\delta_{v(x)} h 
|\varphi_k>
\label{delta-varphi}
,
\ee
and, following from Eq.\ (\ref{h-hf}) 
\bea
\delta_{v(x)} h_{x^\prime x^{\prime\prime}}
&=& 
\sum_{j=1}^N \Big(
\mel{x^{\prime} \,\varphi_j}{\delta_{v(x)}
  V}{x^{\prime\prime}\,\varphi_j}
\nonumber
\\&&+
\mel{x^{\prime} \,\delta_{v(x)} \varphi_j}{V}{x^{\prime\prime}\,\varphi_j}
\nonumber\\
&&+
\mel{x^{\prime}\,\varphi_j}{V}{x^{\prime\prime}\,\delta_{v(x)} \varphi_j}
\Big)
.
\label{delta-h}
\eea
Finally, the 
functional derivative of the orbitals
is obtained
by inserting 
Eq.\ (\ref{delta-h})
into Eq.\ (\ref{delta-varphi})
\begin{eqnarray}
\delta_{v(x)} \varphi_k (x^\prime )
&=&
\sum_{l\ne k} \frac{\varphi_l(x^\prime)}{\epsilon_k-\epsilon_l}
\sum_j^N 
\Big(
\mel{\varphi_l\varphi_j}{\delta_{v(x)} V}{\varphi_k\varphi_j}
\nonumber\\&&+
\mel{\varphi_l\delta_{v(x)}\varphi_j}{V}{\varphi_k\varphi_j}
\nonumber\\&&+
\mel{\varphi_l\varphi_j}{V}{\varphi_k\delta_{v(x)}\varphi_j}
\Big)
\label{invHF}
,
\end{eqnarray}
which
can quite effectively be solved by iteration,
starting for example with initial guess
$\delta_{v(x)}\varphi_j(x^\prime)$ = 0.
Because $\delta_{v(x)}\varphi_k(x^\prime)$
depends on two coordinates $x$ and $x^\prime$,
Eq.\ (\ref{invHF}), being the central equation of the IHFA, 
has the dimension of a two--body equation
for the lowest $N$ orbitals.
Introducing, analogously to $B_i$,
the matrix  $\Delta_{i}(x)$ with  
$\Delta_{kl;i}(x)$ 
= $\scp{x_{i,l}}{\delta_{v(x)} {\varphi_k}}$
= $\delta_{v(x)} {\varphi_k}(x_{i,l})$
(for $x > 0$, $1\le k,l\le N$, $1\le i\le n$),
it is straightforward to check that
the functional derivative of a likelihood term
is given by
\be
\delta_{v(x)} \!\ln p_{\rm HF}(x_i|\hat x,v)
= 2 {\rm Re}\left[{\rm Tr} (B_i^{-1}\Delta_{i}(x))\right]
.
\label{der-likel}
\ee
The stationarity equation
(\ref{posterior-statEQ})
can now be solved by iteration,
for example,
\be
v^{(r+1)}
 \!=
v^{(r)}\! +
\eta \big[ v_0\!-v^{(r)}
\! +{\bf K}_0^{-1}\!\sum_i \delta_x\!\ln p_{\rm HF}(x_i|\hat x,v^{(r)})
\big]
\label{iter1}
,
\ee
choosing 
a positive step width $\eta$
and starting from an initial guess $v^{(0)}$.

In conclusion,
reconstructing a potential from data by IHFA is based on 
the definition of a {\it prior process} for $v$
and requires the iterative solution of

1. the {\it stationarity equation for the potential} (\ref{posterior-statEQ}),
   needing as input for each iteration step (\ref{iter1}) 

2. the {\it functional derivatives of the likelihoods} (\ref{der-likel}),
   obtained by solving the (two--body--like) equation (\ref{invHF})
   for given

3. {\it single particle orbitals}, defined in (\ref{HF-eq}) as solutions of
   the direct (one--body) Hartree--Fock equation (\ref{HF-eq}).

\begin{figure}
\begin{center}
\psfig{file=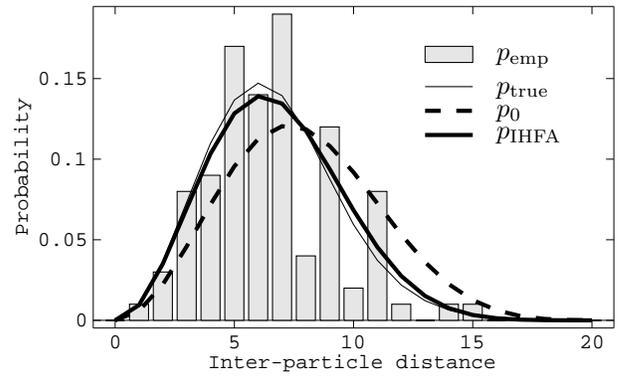, width= 80mm}
\setlength{\unitlength}{1mm}
\begin{picture}(0,0)
\put(-18,42){\makebox(0,0)[l]{$p_{\rm emp}$}}
\put(-18,37.7){\makebox(0,0)[l]{$p_{\rm true}$}}
\put(-18,34.6){\makebox(0,0)[l]{$p_{\rm 0}$}}
\put(-18,31.7){\makebox(0,0)[l]{$p_{\rm IHFA}$}}
\end{picture}
\end{center}
\caption{
Empirical density $p_{\rm emp}$ for $n$ = 100,
true likelihood $p_{\rm true}$, 
reference likelihood $p_0$,     
reconstructed likelihood $p_{\rm IHFA}$
as function of inter--particle distance. 
}
\label{hf-fig1}
\end{figure}

We tested the numerical feasibility
of the IHFA
for a Hamiltonian
$H$ = $-\frac{1}{2m}\Delta+V_1+V$ (with $m$ = $10^{-3}$), 
including a local one--body potential 
with $V_1(z,z^\prime)$ = $\delta(z-z^\prime) a z^2$ (and $a$ = $10^{-5}$)
to break translational symmetry.
We want to reconstruct 
the unknown local two--body potential $V$
from empirical data.
To be able to sample data from
the `true' many--body likelihood (\ref{likelihood})
and to check the quality of the IHFA 
for a given `true' potential $V_{\rm true}$,
we have to solve the corresponding many--body problem.
Therefore, we have chosen a system of two particles 
on a one--dimensional grid (with 21 points) 
for which the true ground state can be calculated 
by diagonalizing $H$ numerically.
We want to stress that application of the
IHFA
to systems with $N>2$ particles is straightforward
and only requires to solve Eq.\ (\ref{invHF}) for $N$
instead 
for two orbitals.
Hence, selecting a `true' local two--body potential 
$V_{\rm true}$
with 
$v_{\rm true}(x)$ = $b/(1+e^{-2 \gamma (x-L/2)/L})$
(and $b$ = 100, $\gamma$ = 10, $L$ = 21),
we were able to sample 100 data points  
from the corresponding `true' probability density (\ref{likelihood}).
The true likelihood 
as function of inter--particle distances $x$ 
and the empirical density of distances
$p_{\rm emp}(x)$ = $(1/n)\sum_{i=1}^n\delta_{x-|x_{i,1}-x_{i,2}|}$
obtained from the training data
are shown in Fig.\ \ref{hf-fig1}.

\begin{figure}
\begin{center}
\epsfig{file=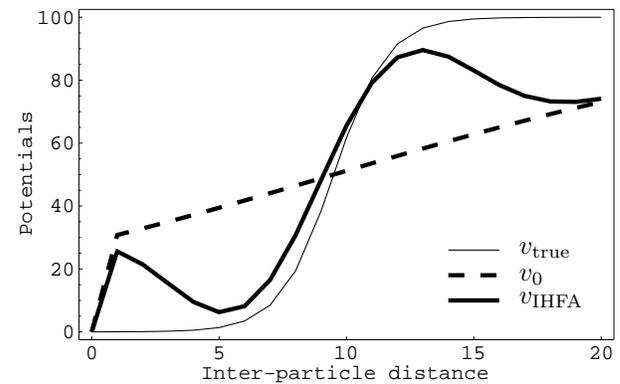, width= 80mm}
\setlength{\unitlength}{1mm}
\begin{picture}(0,0)
\put(-15,18){\makebox(0,0)[l]{$v_{\rm true}$}}
\put(-15,14.5){\makebox(0,0)[l]{$v_0$}}
\put(-15,11.5){\makebox(0,0)[l]{$v_{\rm IHFA}$}}
\end{picture}
\end{center}
\caption{
True potential $v_{\rm true}$,  
reference potential $v_0$, 
and reconstructed IHFA potential $v_{\rm IHFA}$
as function of inter--particle distance. 
}
\label{hf-fig2}
\end{figure}

We used a nonparametric approach for $v$,  %the potential,
combined with a Gaussian smoothness prior  %for $v$,
with inverse covariance ${\bf K}_0$ = $\lambda ({\bf I}-\Delta)/2$
(identity ${\bf I}$, $\lambda$ = $10^{-3}/2$),
and a reference potential 
$v_{0}(x)$ of the form of $v_{\rm true}$, 
but with $\gamma$ = 1 (so it becomes nearly linear in the interval [1,20]).
Furthermore, we have set all potentials to zero at the origin
and constant beyond the right boundary.
The reconstructed potential $v_{\rm IHFA}$
has then been obtained by iterating 
%with ${\bf A}$ = ${\bf K}_0$ 
according to Eq.\ (\ref{iter1}) 
and solving Eqs.\ (\ref{HF-eq}) and (\ref{invHF})
within each iteration step.
The resulting IHFA likelihood 
$p_{\rm IHFA}$ = $p(x|\hat x, v_{\rm IHFA})$
indeed fits well the true likelihood 
$p_{\rm true}$ = $p(x|\hat x, v_{\rm true})$
(see Fig.\ \ref{hf-fig1}).
In particular, $p_{\rm IHFA }$ is over the whole range
an improvement of the reference likelihood 
$p_0$ = $p(x|\hat x, v_{0})$.
The situation is more complex 
for potentials (see Fig.\ \ref{hf-fig2}).
On the basis of 100 data points,
the true potential %$v_{\rm true}$ 
is only well approximated at medium inter--particle distances. 
For large and small distances, on the other hand,
the IHFA solution is still dominated by
the reference potential %$v_0$ 
of the prior process.
This effect is a consequence of the lack of empirical data
in those regions (see Fig.\ \ref{hf-fig1}):
The probability to find particles at large distances is small,
because the true potential has its maximum at large distances.
Also, measuring small distances is unlikely,
because antisymmetry 
forbids two fermions to be at the same place.
In such low data regions one must therefore 
rely on a priori information.

We are grateful to A. Weiguny for stimulating discussions.

\bibliographystyle{prsty}
\bibliography{iqs}

\end{document}